\documentclass[reprint,superscriptaddress,amsmath,amssymb,prl,floatfix]{revtex4-2}
\pdfoutput=1

\usepackage[utf8]{inputenc}
\usepackage{graphicx}
\usepackage{dcolumn}
\usepackage{bm}
\usepackage{hyperref}
\usepackage{mathtools}
\usepackage{xcolor}

\begin{document}

\title{Slowest and Fastest Information Scrambling in the Strongly Disordered XXZ Model}

\author{Myeonghyeon Kim}
\affiliation{Department of Physics and Astronomy, Seoul National University, Seoul 08826, Korea}

\author{Dong-Hee Kim}
\email{dongheekim@gist.ac.kr}
\affiliation{Department of Physics and Photon Science, Gwangju Institute of Science and Technology, Gwangju 61005, Korea}
\affiliation{School of Physics, Korea Institute for Advanced Study, Seoul 02455, Korea}

%\date{\today}

\begin{abstract}
We present a perturbation method to compute the out-of-time-ordered correlator in the strongly disordered Heisenberg XXZ model in the deep many-body localized regime. We characterize the discrete structure of the information propagation across the eigenstates, revealing a highly structured light cone confined by the strictly logarithmic upper and lower bounds representing the slowest and fastest scrambling available in this system. We explain these bounds by deriving the closed-form expression of the effective interaction for the slowest scrambling and by constructing the effective model of a half length for the fastest scrambling. We extend our lowest-order perturbation formulations to the higher dimensions, proposing that the logarithmic upper and lower light cones may persist in a finite two-dimensional system in the limit of strong disorder and weak hopping.
\end{abstract}

\maketitle

Slow scrambling of quantum information is one of the intriguing 
phenomena occurring in many-body localized (MBL) systems
\cite{review2015,review2018a,review2018b,review2019,review2020}. 
The time scale of scrambling dynamics~\cite{Xu2022} 
in MBL systems is distinguished from Anderson localization 
in noninteracting systems where correlation decays exponentially 
\cite{Burrell2007,Hamza2012}
and also from the fast scrambling expected in ideal chaotic systems
\cite{Maldacena2016,Roberts2016,Gu2017}.
The logarithmic time scale of information propagation was 
first reported by the growth of entanglement entropy 
in the disordered XXZ chain quenched from a product state
\cite{Chiara2006,Znidaric2008,Bardarson2012},
which was explained in the picture of 
the quasilocal integral of motion (LIOM) 
\cite{Serbyn2013a,Vosk2013,Serbyn2013b,Huse2014}.
The Lieb-Robinson bound indicating the upper bound on 
information propagation speed was modified accordingly
in this picture, 
proposing the logarithmic light cone (LLC) of 
the information front moving at a finite speed defined 
in logarithmic time instead of linear time
\cite{Kim2014,Fan2017,Swingle2017,ChenY2017,ChenX2017,He2017}.

Despite the numerical evidence of LLC found in MBL systems
\cite{Deng2017,He2017,Huang2017,Banuls2017,Tomasi2019,Kim2021},
a basic understanding of LLC primarily relies on
the effective l-bit Hamiltonian \cite{Serbyn2013b,Huse2014}.
The hypothesized exponentially decaying effective interaction
$J_\mathrm{eff}(r) \propto \exp(-r/\xi)$ 
acting on two remote LIOMs at distance $r$ with a decay length $\xi$
is a key to interpreting 
the time scale $t \sim 1/J_\mathrm{eff}$ exponentially increasing with $r$. 
Although this is well established to describe the dephasing dynamics 
in one dimension (1D), the effective picture lacks 
the system-specific details that can still be necessary for understanding 
of the phenomena in a particular system.
In the simple setting with a fixed $\xi$, the slope of LLC is given 
by $\xi^{-1}$ \cite{Fan2017}.
However, as noted in the construction of the l-bit model \cite{Huse2014},
$J_\mathrm{eff}$ and $\xi$ generally vary with eigenstates as well as 
disorder configurations. We study the consequence of such dependence 
in characterizing information scrambling 
in the disordered XXZ model in the deep MBL regime.

On the other hand, practical signatures of MBL 
in two dimensions (2D) have attracted much attention theoretically 
\cite{Wahl2019,Kshetrimayum2020,Theveniaut2020,Szabo2020,Chertkov2021,Tomasi2019,Kim2021,Pietracaprina2021,Decker2022,Doggen2022,Agrawal2022,Strkalj2022,Tang2021,Li2021,Foo2022,Venn2022}
and experimentally \cite{Choi2016,Kondov2015,Bordia2017}
at finite systems,
while it has been argued that 2D MBL is asymptotically unstable
toward the avalanche of rare thermal regions
\cite{DeRoeck2017,DeRoeck2017review,Potirniche2019,Gopalakrishnan2019,Doggen2020}.
In particular, the evidence of LIOMs \cite{Chertkov2021} and 
LLC \cite{Kim2021} has been recently presented 
in higher dimensions by the numerical construction of the l-bit Hamiltonian.
These motivate us to revisit the computation of 
the out-of-time-ordered correlator (OTOC) \cite{Xu2022}, 
a diagnostic tool for information scrambling, 
for characterization beyond the generic l-bit description both in 1D and 2D.

\begin{figure}[b]
    \centering
    \includegraphics[width = 0.48\textwidth]{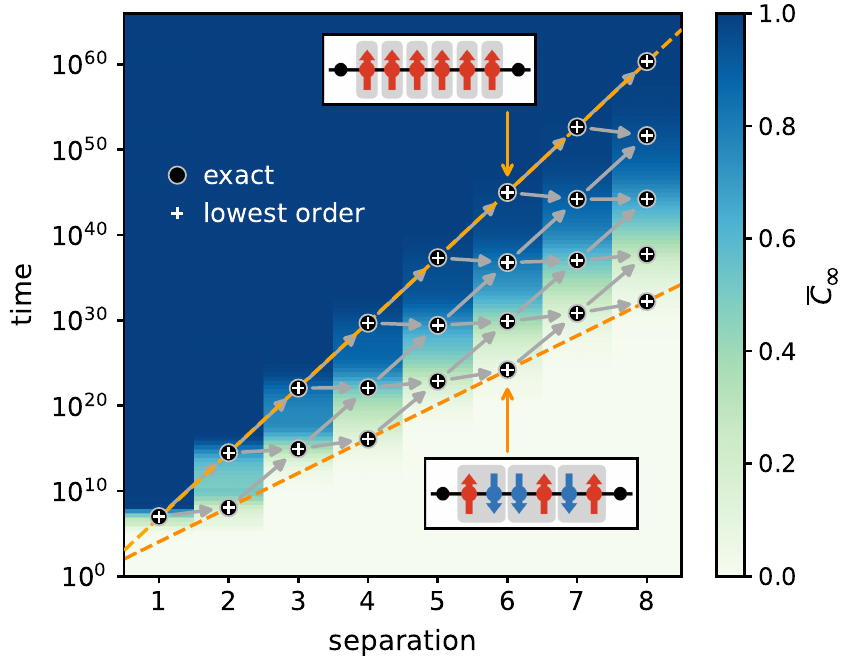}
    \caption{Light cone structure of the disordered XXZ chain
    in the deep MBL regime. 
    The markers denote the scrambling time $J_z t^*$ obtained at the fixed value 
    of the disorder-averaged OTOC 
    $\langle C_\alpha(r,t^*) \rangle_\mathrm{av} = 0.5$ 
    for each eigenstate $\alpha$, 
    comparing the lowest-order perturbation results 
    with the exact diagonalization at $J/J_z = 0.001$ and $h/J_z = 10$. 
    The arrows indicate the allowed change of $t^*$ with increasing separation. 
    The background color indicates the infinite-temperature OTOC
    $\overline{C}_\infty$, the average of  
    $\langle C_\alpha(r,t) \rangle_\mathrm{av}$ over all eigenstates.
    }
    \label{fig1}
\end{figure}

In this Letter, we develop a perturbation formulation of OTOC 
in the strongly disordered XXZ model in the weak hopping limit. 
Measuring OTOC for each eigenstate, we reveal the discrete structure of 
the light cone built by the allowed lowest orders of perturbation 
varying with the intervening spin states at a given $r$. 
Remarkably, the light cone is bounded by the two logarithmic slopes 
representing the slowest and fastest scrambling. 
We derive an analytic formula for the effective interaction for 
the slowest scrambling and describe the fastest scrambling by 
the half length effective Ising chain. Extending our method to 2D, 
we demonstrate the logarithmic light cones of the slowest and fastest scrambling 
in 2D within the lowest-order perturbation formulations.

For perturbation expansion, we decompose the XXZ Hamiltonian 
as $\hat{H} \equiv \hat{H}_0 + \hat{V}$,
where the unperturbed part $\hat{H}_0$ and  
the hopping perturbation $\hat{V}$ are given as
\begin{eqnarray}
    \hat{H}_0 &=& \frac{J_z}{2} \sum_i \hat{\sigma}^z_i \hat{\sigma}^z_{i+1} 
    + \sum_i h_i \hat{\sigma}^z_i ,\\
    \hat{V} &=& J \sum_i \left( \hat{\sigma}^+_i \hat{\sigma}^-_{i+1} 
    + \hat{\sigma}^-_i \hat{\sigma}^+_{i+1} \right) .
\end{eqnarray}
The random disorder field is drawn 
from the uniform distribution of $h_i \in [-h, h]$. 
We assume that the unperturbed state is nondegenerate
and localized in the Fock space of the $\hat{\sigma}_z$-basis states. 
We consider the strong disorder and weak hopping limit of 
$J \ll J_z \ll h$ in the deep MBL regime. 
We compute the perturbation corrections in energy
within the Rayleigh-Schr\"odinger perturbation theory
using multiprecision numerics to handle strong cancellations
and critical round-off errors (see Supplemental Material \cite{sm}
and references \cite{Kato1949,Bloch1958,Huby1961,Silverstone1970} therein). 

We define OTOC by the squared commutator of 
two $\hat{\sigma}_x$ operators initially located at $a$ and $b$ as
\begin{equation} \label{eq:otoc_def}
    C_\alpha(r,t) = \frac{1}{2} \langle \alpha |
    \left|[\hat{\sigma}^x_a(t), \hat{\sigma}^x_b]\right|^2 
    | \alpha \rangle 
    = 1 - \mathrm{Re} [F_\alpha(r, t)], 
\end{equation}
where the correlator 
$F_\alpha(r, t) = \langle \alpha | \hat{\sigma}^x_a(t) \hat{\sigma}^x_b \hat{\sigma}^x_a(t) \hat{\sigma}^x_b | \alpha \rangle$,
and $r \equiv |a - b| - 1 \ge 0$ is the separation between $a$ and $b$.
Choosing $|\alpha\rangle$ to be an eigenstate,
the correlator can be approximated at weak perturbation as 
\begin{equation} \label{eq:otoc_approx}
    F_\alpha(r,t) = \sum_{\beta,\gamma,\delta} 
    s_{\alpha\beta\gamma\delta} 
    e^{i\Omega_{\alpha\beta\gamma\delta}t}
    \approx \exp(iJ_\mathrm{eff}^\alpha t),
\end{equation}
where the frequency 
$\Omega_{\alpha\beta\gamma\delta} =  E_\alpha - E_\beta + E_\gamma - E_\delta$ 
and the coefficient 
$s_{\alpha\beta\gamma\delta} = 
\langle \alpha| \hat{\sigma}^x_a |\beta \rangle
\langle \beta| \hat{\sigma}^x_b |\gamma \rangle
\langle \gamma| \hat{\sigma}^x_a |\delta \rangle
\langle \delta| \hat{\sigma}^x_b |\alpha \rangle$.
Assuming that a perturbation correction
in a state vector is small, the single dominant term 
is found at $s_{\alpha\beta\gamma\delta} \approx 1$ for 
$|\alpha\rangle\ \approx |\alpha^{(0)}\rangle$,
$|\beta\rangle \approx |\beta^{(0)}\rangle = \hat{\sigma}^x_a |\alpha^{(0)}\rangle$,
$|\gamma\rangle \approx |\gamma^{(0)}\rangle = \hat{\sigma}^x_b \hat{\sigma}^x_a |\alpha^{(0)}\rangle$, and 
$|\delta\rangle \approx |\delta^{(0)}\rangle = \hat{\sigma}^x_b |\alpha^{(0)}\rangle$, 
where the superscript denotes the corresponding unperturbed state.
The frequency of the dominant component is rewritten 
in terms of the perturbation corrections as
\begin{equation} \label{eq:Jeff_pt}
    J_\mathrm{eff}^\alpha = \Delta E_\alpha - \Delta E_\beta 
    + \Delta E_\gamma - \Delta E_\delta,
\end{equation}
which we referred to as an effective interaction from
the analogy to the one in $F(t) = \exp(\pm 4i J_\mathrm{eff} t)$ 
given for the effective l-bit model \cite{Fan2017,Swingle2017,ChenY2017,ChenX2017,He2017}.
The same expression of $J^\alpha_\mathrm{eff}$ can also be 
extracted using the protocol of the double electron-electron 
resonance (DEER) \cite{Serbyn2014,Kucsko2018,Varma2019,Chiaro2022}.
From Eqs. \eqref{eq:otoc_def} and \eqref{eq:otoc_approx}, 
the disorder average of $C_\alpha$ is written as 
\begin{equation} \label{eq:OTOCav}
    \left\langle C_\alpha (r,t) \right\rangle_\mathrm{av} 
    \approx 1 - \mathrm{Re}
    \left[ 
    \int^\infty_{-\infty} e^{iJ^\alpha_\mathrm{eff}t} P(J^\alpha_\mathrm{eff})\,\mathrm{d}J^\alpha_\mathrm{eff}
    \right]
\end{equation}
with the probability distribution $P(J^\alpha_\mathrm{eff})$ being obtained
by computing $J^\alpha_\mathrm{eff}$ for random disorder configurations.
In this weak perturbation formulation, only the energy corrections 
are important while the small corrections in the state vectors are 
irrelevant. Measuring OTOC in the Fock space with
$|\alpha^{(0)}\rangle$ leads to the same expression.

\begin{figure}
    \centering
    \includegraphics[width = 0.48\textwidth]{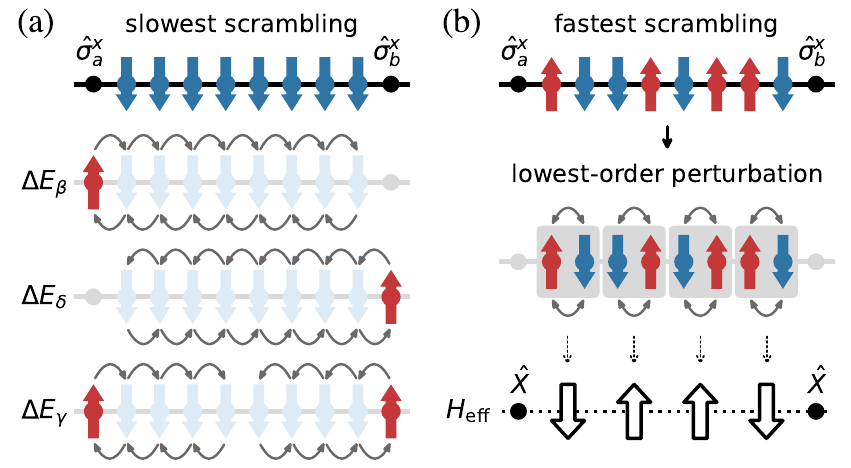}
    \caption{Schematic diagrams of the lowest-order contributions 
    to the effective interactions. 
    (a) In the slowest scrambling, the lowest order is given by
    the minimum sequential moves of the excitation covering 
    the intervening spin-polarized region.
    (b) In the fastest scrambling across the blocks of staggered spin pairs, 
    the lowest order only involves spin exchanges within the block,
    mapping the block into one Ising pseudospin.
    }
    \label{fig2}
\end{figure}

Figure~\ref{fig1} displays the scrambling time $t^*$ 
as a function of $r$ obtained by solving 
$\langle C_\alpha(r,t^*)\rangle_\mathrm{av} = 0.5$ 
for each eigenstate. It turns out that $t^*$ is not on a single light cone
but structured by the lowest order of the nonvanishing perturbation term 
in Eq.~\eqref{eq:Jeff_pt}, varying with
the intervening spin configuration in $|\alpha^{(0)}\rangle$.
The lowest order $n_\alpha(r)$ is determined by the minimum number 
of the hopping operators flipping all intervening spins, which is written as 
$n_\alpha(r) = 2 ( r - m^\alpha_s)$, where 
$m^\alpha_s$ is the number of staggered spin pairs
found in $|\alpha^{(0)}\rangle$ between $a$ and $b$. 

Remarkably, the discrete structure of $t^*$ indicates the
sharp upper and lower bounds in the logarithmic slope, 
representing the slowest and fastest scrambling available
in this system. 
These bounds correspond to $J_\mathrm{eff}^\alpha \propto J^{2r}$ 
($m^\alpha_s = 0$) and $J_\mathrm{eff}^\alpha \propto J^r$ 
($m^\alpha_s = r/2$) at even $r$, which are
associated with $|\alpha^{(0)}\rangle$ of the ferromagnetic (FM)
domain and the chain of staggered spin pairs 
such as in the antiferromagnetic (AF) state, respectively.
This structure is hidden in the infinite-temperature
OTOC, an average over the eigenstates, revealing 
the detailed view of the light cone in the 
strongly disordered XXZ model.

For the slowest scrambling, we obtain the lowest-order expression of 
$J_\mathrm{eff}^\mathrm{FM}$ at the FM unperturbed state as
\begin{equation} \label{eq:Jeff_FM}
    J_\mathrm{eff}^\mathrm{FM} =  2 J_z \left(\frac{J}{2}\right)^{2r}
    \sum_{k=0}^{r} F_k^2 \frac{A_k + B_{k+1}}{A_k + B_{k+1} - J_z} G_{k+1}^2,
\end{equation}
where $A_k = h_{a} - h_{a+k}$ and $B_k = h_{b} - h_{a+k}$. 
The factors are given as $F_k = \prod_{j=1}^k A_j^{-1}$ 
and $G_k = \prod_{j = k}^r B_j^{-1}$, where an empty product is unity.
Note that a nonzero interaction $J_z$ is essential.
While the detailed derivation is provided in the Supplemental 
Material \cite{sm}, each term is conceptually illustrated 
in Fig.~\ref{fig2}(a). 
The diagrams of $\Delta E_\beta$ and $\Delta E_\delta$
describe the lowest order of $\hat{V}$ that moves an excitation
site by site to sweep through the intervening FM area and
$\Delta E_\gamma$ includes all such hopping configurations 
with two excitations.

\begin{figure}
    \centering
    \includegraphics[width = 0.48\textwidth]{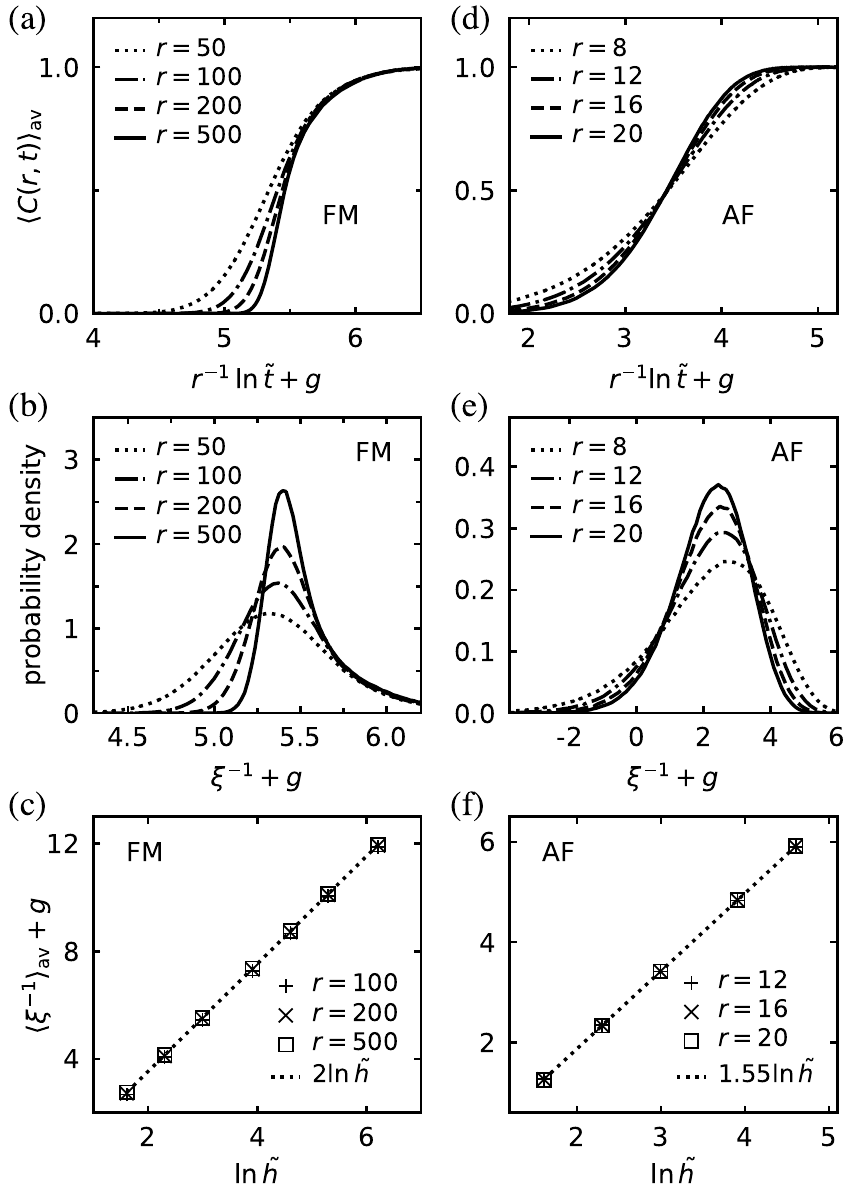}
    \caption{Slowest and fastest scrambling in the lowest-order
    perturbation theory for the strongly disordered XXZ chain.
    The disorder average $\langle C(r,t) \rangle_\mathrm{av}$, 
    the distribution of $\xi^{-1} \equiv -r^{-1}\ln|\tilde{J}_\mathrm{eff}|$, 
    and the decay length $\langle \xi^{-1} \rangle_\mathrm{av}$ are computed 
    at $h/J_z = 20$ 
    for the (a)--(c) FM and (d)--(f) AF states.
    The constant $g$ is set to be $2 \ln \tilde{J}$ (FM) and $\ln \tilde{J}$ (AF). 
    }
    \label{fig3}
\end{figure}

The fastest scrambling in the lowest-order picture is described by
a half number of pseudospins each of which maps to a two-site block
of a staggered spin pair as sketched in Fig.~\ref{fig2}(b).
The lowest order is given by the $r/2$ number of the $\hat{V}$ operators
applying exclusively on each block for the simultaneous flip of 
the two opposite spins.
The resulting two-level structure leads us to define the pseudospin 
Pauli operators $\hat{X}$ and $\hat{Z}$ 
in the basis of $|\Uparrow\rangle \equiv |\downarrow \uparrow \rangle$ 
and $|\Downarrow\rangle \equiv |\uparrow \downarrow \rangle$ 
for the reduced Hilbert space. 
We choose the AF state to evaluate $J_\mathrm{eff}^\mathrm{AF}$, 
but all configurations filled up with staggered spin pairs 
provide the equivalent results.

At the lowest order, $J_\mathrm{eff}^\mathrm{AF}$ in
the XXZ chain is exactly reproduced by the Ising chain 
of a half length $l = r/2$,
\begin{equation} \label{eq:Heff}
    \hat{H}_\mathrm{Ising} = 
    -\frac{J_z}{2} \sum_{k=0}^l \hat{Z}_k \hat{Z}_{k+1} 
    + \sum_{k=0}^{l+1} \Delta_k \hat{Z}_k 
    + J \sum_{k=1}^l \hat{X}_k\,, 
\end{equation}
where $\Delta_0 = h_a$, $\Delta_{l+1} = -h_b$, and
$\Delta_k = h_{a+2k} - h_{a + 2k -1}$ for $k = 1,\ldots,l$. 
The perturbation part is $J\sum_i \hat{X}_i$. The FM state
corresponds to the AF state of the XXZ chain and
$\hat{X}_{0,l+1}$ replaces $\hat{\sigma}^x_{a,b}$ 
for the OTOC operators.
While we cannot find an analytic formula of $J_\mathrm{eff}^\mathrm{AF}$,
the half length chain significantly reduces 
the numerical cost for the full perturbation calculation \cite{sm}. 
Since nonzero $J_z$ is essential in both 
$J_\mathrm{eff}^\mathrm{FM}$ and $J_\mathrm{eff}^\mathrm{AF}$,
hereafter we express the quantities in a dimensionless form as
$\tilde{t} \equiv J_z t$, $\tilde{h} \equiv h/J_z$, 
$\tilde{J} \equiv J/J_z$, and $\tilde{J}_\mathrm{eff} \equiv J_\mathrm{eff}/J_z$.

Figure~\ref{fig3} presents the numerical results
based on Eqs. \eqref{eq:Jeff_FM} and \eqref{eq:Heff},  
which verifies the logarithmic propagation of
the fronts of the slowest and fastest scrambling
but also examines the decay length scale of the effective interaction.
The disorder-averaged OTOC plotted as a function of 
$r^{-1}\ln \tilde{t}$ exhibits an increase that gets sharper 
as $r$ increases, assuring the strictly logarithmic slopes of 
the light cone. The shift $g \equiv q\ln\tilde{J}$ comes from 
$\tilde{J}_\mathrm{eff} \propto \tilde{J}^{qr}$ where $q=2(1)$ 
is for the FM(AF) state.

Assuming the form of $\tilde{J}_\mathrm{eff} \sim \exp(-r/\xi)$,
we extract the inverse decay length as 
$\xi^{-1} = - r^{-1}\ln|\tilde{J}_\mathrm{eff}|$. 
The distribution of $\xi^{-1}$ is increasingly peaked as $r$ increases,
indicating a well-defined $\langle \xi^{-1} \rangle_\mathrm{av}$. 
The skewed shape that we observe here at the particular states
is different from the log-normal shape previously 
reported at infinite temperature \cite{Varma2019}.
In addition, we find that $\langle \xi^{-1} \rangle_\mathrm{av}$ 
follows the characteristic behavior with varying parameters as
\begin{equation} \label{eq:decay_length}
    \left\langle \xi^{-1} \right\rangle_\mathrm{av}  = 
    - \frac{\langle \ln|\tilde{J}_\mathrm{eff}| \rangle_\mathrm{av}}{r}
    \sim 
    \begin{cases}
        \ln(\tilde{h}/\tilde{J})^2 & \text{for FM}, \\
        \ln (\tilde{h}^\kappa/\tilde{J}) & \text{for AF}.
    \end{cases}
\end{equation}
One can directly extract the behavior for the FM state from 
Eq.~\eqref{eq:Jeff_FM} giving
$\tilde{J}_\mathrm{eff}^\mathrm{FM} \sim (\tilde{J}/2\tilde{h})^{2r}$
after rewriting it in the dimensionless form. For the AF state,
we determine the exponent $\kappa \approx 1.55$ numerically.

\begin{figure*}
    \centering
    \includegraphics[width = 0.96\textwidth]{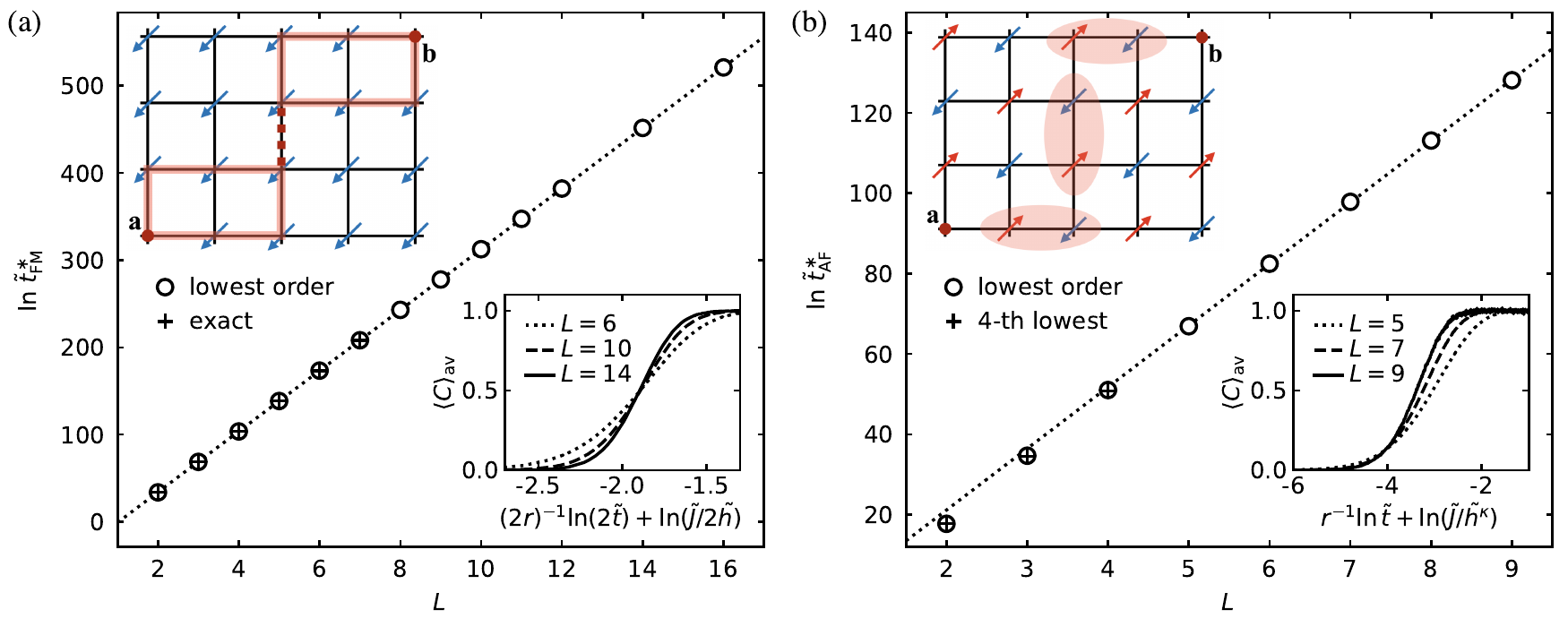}
    \caption{Logarithmic light cones in the 2D strongly disordered 
    XXZ model. The scrambling time $t^*$ is computed 
    at $J/J_z = 0.001$ and $h/J_z = 20$ for the (a) FM and (b) AF states, 
    corresponding to the slowest and fastest scrambling, respectively.
    In the $(L+1) \times L$ lattices,
    the $\hat{\sigma}_x$ operators of OTOC are located 
    at the diagonal corners with separation $r = 2 (L - 1)$.
    The schematic diagram in the insets shows an example of 
    a path contributing to the lowest-order perturbation calculation.
    }
    \label{fig4}
\end{figure*}

Our lowest-order formulations developed above in 1D can be readily
extended to 2D by considering the multiple paths 
of the same Manhattan distance between the two sites $\mathbf{a}$ and $\mathbf{b}$,
namely the number of edges to hop along the path,
composing the non-vanishing lowest-order terms.
Below we describe the calculations of $J_\mathrm{eff}$ 
at the FM and AF states in $L_x \times L_y$ lattices 
with the two operators being located at  
the opposite corners as sketched in Fig.~\ref{fig4}. 
We remove boundary artifacts by adding the FM or AF
environments to the system.

For the FM state, the lowest order is determined as
$2r(\mathbf{a},\mathbf{b}) = 2(L_x + L_y - 3)$, which depends on the number 
of sites along the shortest paths between $\mathbf{a}$ and $\mathbf{b}$.
The 2D variant of Eq.~\eqref{eq:Jeff_FM}
is written in a dimensionless form as 
\begin{equation} \label{eq:Jeff_2DFM}
    \tilde{J}_\mathrm{eff}^\mathrm{FM} = 
    2 \left(\frac{\tilde{J}}{2\tilde{h}}\right)^{2r}
    \sideset{}{^\prime}{\sum}_{\mathclap{(\mathbf{x}_1\to\mathbf{x}_2)}}
    \tilde{F}_{\mathbf{x}_1}^2
    \frac{\tilde{A}_{\mathbf{x}_1} + \tilde{B}_{\mathbf{x}_2}}{\tilde{A}_{\mathbf{x}_1} + \tilde{B}_{\mathbf{x}_2} - \tilde{h}^{-1}}
    \tilde{G}_{\mathbf{x}_2}^2, 
\end{equation}
where 
$\tilde{A}_\mathbf{x} = (\tilde{h}_\mathbf{a} - \tilde{h}_\mathbf{x})/\tilde{h}$ and
$\tilde{B}_\mathbf{x} = (\tilde{h}_\mathbf{b} - \tilde{h}_\mathbf{x})/\tilde{h}$.
The primed sum runs over directed links $(\mathbf{x}_1 \to \mathbf{x}_2)$ 
on any shortest path from $\mathbf{a}$ to $\mathbf{b}$.
The factors $\tilde{F}$ and $\tilde{G}$ are defined as
\begin{equation*}
    \tilde{F}_\mathbf{x} = \sum_{w(\mathbf{a},\mathbf{x})}
    \sideset{}{^\mathbf{a}}{\prod}_{\mathbf{y} \in w} \tilde{A}_\mathbf{y}^{-1}, 
    \quad
    \tilde{G}_\mathbf{x} = \sum_{w(\mathbf{b},\mathbf{x})} 
    \sideset{}{^\mathbf{b}}\prod_{\mathbf{y} \in w} \tilde{B}_\mathbf{y}^{-1},
\end{equation*}
where the sum runs over every shortest path $w(\mathbf{x}_0, \mathbf{x})$ 
connecting $\mathbf{x}_0$ and $\mathbf{x}$ and  
$\prod^{\mathbf{a}(\mathbf{b})}$ excludes $\mathbf{a}$($\mathbf{b}$) 
in the product over every site $\mathbf{y}$ along the path $w$. The squared
factors consider the excitation moving forward 
and backward along different paths unlike in 1D.

For the AF state, we consider $(L+1) \times L$ lattices,
where $l \equiv L-1$ pairs of the up-and-down spins exist along any shortest path 
between $\mathbf{a}$ and $\mathbf{b}$, giving the lowest order $r = 2l$.
Unlike the FM case, the lowest-order contributions can be separated
into each path because a string of the hopping 
operators for paired spin flips must stay on the same path. 
For a path
$w \equiv (\mathbf{a}, \mathbf{x}_1, \mathbf{x}_2, \ldots, \mathbf{x}_{2l}, \mathbf{b})$,
the contribution is then given by 
the Ising chain with path-dependent parameters,
which can be expressed as
$\hat{H}_\mathrm{Ising}[\boldsymbol{\Delta}(w)]$
with $\Delta_0 = h_\mathbf{a}$, $\Delta_{l+1} = - h_\mathbf{b}$, and
$\Delta_k = h_{\mathbf{x}_{2k}} - h_{\mathbf{x}_{2k-1}} + 2J$,
where $2J$ is from the AF surroundings. 
Summing over all paths, we write $\tilde{J}_\mathrm{eff}^\mathrm{AF}$ as
\begin{equation} \label{eq:Jeff_2DAF}
    \tilde{J}_\mathrm{eff}^\mathrm{AF} (\mathbf{a}, \mathbf{b}) = \sum_w 
    \tilde{J}_\mathrm{eff}^\mathrm{AF}\bm{[}\hat{H}_\mathrm{Ising}[\boldsymbol{\Delta}(w)]\bm{]},
\end{equation}
which involves an exponentially growing number of terms as $L$
increases but allows us to go well beyond the system-size limit of
the exact diagonalization and the numerical perturbation
calculations for arbitrary orders.

Figure~\ref{fig4} shows 2D LLCs from the scrambling time and 
the disorder-averaged OTOC measured at the FM and AF states
in the 2D XXZ model in the strong disorder and weak hopping limit. 
Since the number of the shortest paths scales as $4^{l}$,
a rough estimate ignoring disorder correlations between the paths
suggests $\tilde{J}_\mathrm{eff}^\mathrm{AF} \sim 4^l e^{-2l/\xi}$
from Eq.~\eqref{eq:Jeff_2DAF}, implying LLC for 
$\langle \xi^{-1} \rangle_\mathrm{av} \gg \ln 2$.
While our calculations are based on the lowest-order perturbation theory, 
the numerical tests show excellent agreement with
the exact diagonalization at small $L$'s for
the FM state and with the full perturbation calculations
up to the fourth lowest order for the AF state.
Our observation of LLC in the strongly disordered XXZ model 
is also consistent with the previous evidence
of LLC reported in the 2D bosonic system with 
the l-bit construction at the strong disorder 
and weak interaction limit \cite{Kim2021}. 

In conclusion, our perturbation formulation reveals 
the peculiar structure of slow information propagation 
in the paradigmatic XXZ model in the deep MBL regime. 
The slowest and fastest scrambling identified in 
the discrete structure of OTOC characterizes 
the drastic difference between the spin-polarized and 
the N\'eel states of the intervening spins prepared 
for the OTOC or DEER measurements. 
We have derived the closed-form expression of the effective interaction 
for the slowest scrambling and found the effective Ising chain 
of a half length describing the fastest scrambling, 
presenting the sharp logarithmic upper and lower bounds 
of the light cone.

Our observation of LLCs extends the variety of the practical 
MBL signatures previously reported in finite 2D systems, 
although the instability of 2D MBL in the asymptotic limit 
goes beyond our method. A challenging direction
for future study may include the behavior of OTOC measured 
across 2D thermal defects and its finite-size effects.
On the other hand, our findings on the distance effectively reduced 
by half at the fastest scrambling imply an interesting question 
on its l-bit representation.
In contrast to the slowest one, the fastest scrambling involves
only the half number of the pseudospins, proposing to further explore 
how the mapping to the l-bit Hamiltonian encodes
these system-specific scrambling structures for the XXZ model.

\begin{acknowledgments}
This work was supported from the Basic Science Research Program 
through the National Research Foundation of Korea (Grant No. NRF-2017R1D1A1B03034669) 
and also from a GIST Research Institute (GRI) grant funded by the GIST.
We appreciate APCTP for its hospitality during the completion of this work.
\end{acknowledgments}

\bibliography{paper}

\end{document}

\title{Supplemental Material for ``Slowest and Fastest Information Scrambling in the Strongly Disordered XXZ Model''}

\author{Myeonghyeon Kim}
\affiliation{Department of Physics and Astronomy, Seoul National University, Seoul 08826, Korea}

\author{Dong-Hee Kim}
\email{dongheekim@gist.ac.kr}
\affiliation{Department of Physics and Photon Science, Gwangju Institute of Science and Technology, Gwangju 61005, Korea}
\affiliation{School of Physics, Korea Institute for Advanced Study, Seoul 02455, Korea}

\maketitle

\section{Numerical arbitrary-order perturbation calculations}

The energy correction of an arbitrary order 
in the Rayleigh-Schr\"odinger perturbation theory (RSPT)
appears in various forms in the literature \cite{Kato1949,Bloch1958,Huby1961} 
(see also references in \cite{Silverstone1970}).
For our case with zero diagonals in the perturbation matrix, 
only the even-order terms survive in RSPT, which can be written 
in a recursive form as
\begin{equation} \label{eq:RSPT}
    E^{(2n)}_\lambda = \big\langle \lambda \big| \hat{V} \big( \hat{P}_\lambda\hat{V} \big)^{2n-1} \big| \lambda \big\rangle 
    + \sum^{n-1}_{m=1} (-1)^m \sum^{n-1}_{k=m}
    \left[\,\, 
    \sideset{}{'}\sum_{(z_1, \ldots, z_m)} \prod^m_{i=1} E^{(2z_i)}_\lambda 
    \right]
    \left[\,\,\,
    \sideset{}{''}\sum_{\substack{(z_1, \ldots, z_q) \\ q \equiv 2n-2k-1}}
    \big\langle \lambda \big| \hat{V} \prod_{i=1}^q \big[ \hat{P}_\lambda^{1+z_i} \hat{V} \big] \big| \lambda \big\rangle
    \right]\,,
\end{equation}
where $|\lambda\rangle$ denotes an unperturbed eigenstate, and the operator $\hat{P}_\lambda = \sum_{\lambda' \ne \lambda}\frac{|\lambda'\rangle\langle\lambda'|}{E^{(0)}_\lambda - E^{(0)}_{\lambda'}}$. 
The primed sum $\sum^\prime$ indicates
the sum over the sequences of $m$ natural numbers $(z_1, z_2, \ldots, z_m)$ 
satisfying the constraint $\sum_{i=1}^m z_i = k$.
The double-primed sum $\sum^{\prime\prime}$ indicates the sum over
the sequences of $q \equiv 2n-2k-1$ non-negative integers 
$(z_1, z_2, \ldots, z_q)$ satisfying the constraint $\sum_{i=1}^q z_q = m$.

\section{Numerical calculation details}

Let us begin with a few remarks on the numerical implementation of Eq.~\eqref{eq:RSPT}. 
First, the number of terms to be computed increases combinatorially
as it goes to the higher order, which practically limits 
the order that one can reach within a reasonable computational time.
Second, round-off errors are often dangerous when computing  
$\Delta E_\alpha - \Delta E_\beta + \Delta E_\gamma - \Delta E_\delta$ 
since the difference between those energy corrections
is much smaller than the magnitude of $\Delta E$.
The difference decreases exponentially with increasing $r$ at the lowest order
in the deep MBL regime that we consider, and thus extremely accurate
calculations are necessary to cope with strong cancellations occurring 
in the perturbation expansion.
To prevent these issues, our numerical calculations keep at least $500$ 
decimal digits by using the multiprecision \texttt{MPFR} library \cite{MPFR}.
Third, the perturbation matrix $\langle \lambda | \hat{V} | \lambda' \rangle$ 
has a highly sparse structure, which helps us to accelerate 
the evaluation of Eq.~\eqref{eq:RSPT} 
by using fast sparse matrix-vector multiplications.

The use of the \texttt{MPFR} library is not limited to the evaluation 
of Eq.~\eqref{eq:RSPT}. The numerical precision of $500$ decimal digits
is applied to all calculations including the exact diagonalization and
the numerical evaluations of $J_\mathrm{eff}$ for the slowest
and fastest scrambling to prevent the same accuracy issues.

The random disorder configurations are generated using the 64-bit version of 
the Mersenne Twister pseudorandom number generator \cite{MT19937} and 
its multiprecision extension implemented in the \texttt{boost} library. 
The number of disorder configurations used for statistics 
in our numerical calculations is listed as follows.
In Fig.~1 and Fig.~3 of the main paper, we use $10^4$ disorder configurations 
for the exact diagonalization calculations and $10^6$ configurations
for the perturbation calculations.
In Fig.~4, for the 2D FM state, the number of disorder configurations used in
the lowest-order perturbation calculations is $10^6$ for $L \le 12$, 
$5\times 10^5$ for $L=14$, and $5\times 10^4$ for $L=16$.
The exact diagonalization results are obtained with $10^4$ configurations.
For the 2D AF state, the number of disorder configurations used 
in the lowest-order calculations is $10^6$ for $L \le 6$, 
$2\times 10^5$ for $L = 7$, $10^5$ for $L=8$, and $10^4$ for $L = 9$
while the higher-order full RSPT calculations are done for $10^4$ configurations.

\section{Derivation of the lowest-order term for the slowest scrambling}

In this section, we focus on the ferromagnetic unperturbed eigenstate 
$|\alpha \rangle = |\cdots00_00_10_2\cdots0_r0_{r+1}0\cdots\rangle$ 
corresponding to the upper light cone of the slowest scrambling, 
where $|0\rangle$ and $|1\rangle$ denote $|\downarrow\rangle$ and 
$|\uparrow\rangle$ in the $\sigma^z$-basis, respectively.
The subscript denotes the site index of the intervening spin 
segment in the XXZ chain, where $0$ and $r+1$ are the locations of 
$\hat{\sigma}^x_a$ and $\hat{\sigma}^x_b$ operators defining OTOC. 
For this particular state with the spin-polarized domain, it is possible 
to derive the closed-form lowest-order expression of the effective interaction
$J_\mathrm{eff}^\mathrm{FM} \equiv \Delta E_\alpha - \Delta E_\beta + \Delta E_\gamma - \Delta E_\delta$ 
by diagrammatically evaluating the energy correction terms in RSPT \cite{SzaboBook,ShavittBook}. 
Below we present our derivation of the lowest $(2r)^\mathrm{th}$ order perturbation
correction of $E_\beta$, $E_\gamma$, and $E_\delta$. Note that 
$\Delta E_\alpha = 0$ because $\hat{V}|\alpha\rangle$ = 0.
For a more intuitive description, let us call the spin excitation $|1\rangle$ just as a particle.

\subsection{single excitation}

The unperturbed eigenstates $|\beta\rangle$ and $|\delta\rangle$ 
with single excitation from $|\alpha\rangle$ created by $\hat{\sigma}^x_0$ 
and $\hat{\sigma}^x_{r+1}$ are given as 
\begin{eqnarray}
|\beta\rangle &=& \hat{\sigma}^x_0 |\alpha \rangle = |1_0 0_1 0_2 \cdots 0_{r-1} 0_r 0_{r+1}\rangle\,, \\
|\delta\rangle &=& \hat{\sigma}^x_{r+1} |\alpha \rangle = |0_0 0_1 0_2 \cdots 0_{r-1} 0_r 1_{r+1}\rangle\,,
\end{eqnarray}
where the configuration in the outside of the region $[0, r+1]$ is omitted 
because it is irrelevant to the lowest-order perturbation expression of 
$J_\mathrm{eff}^\alpha \equiv \Delta E_\alpha - \Delta E_\beta + \Delta E_\gamma - \Delta E_\delta$. 

\begin{figure}
    \centering
    \includegraphics[width=0.85\textwidth]{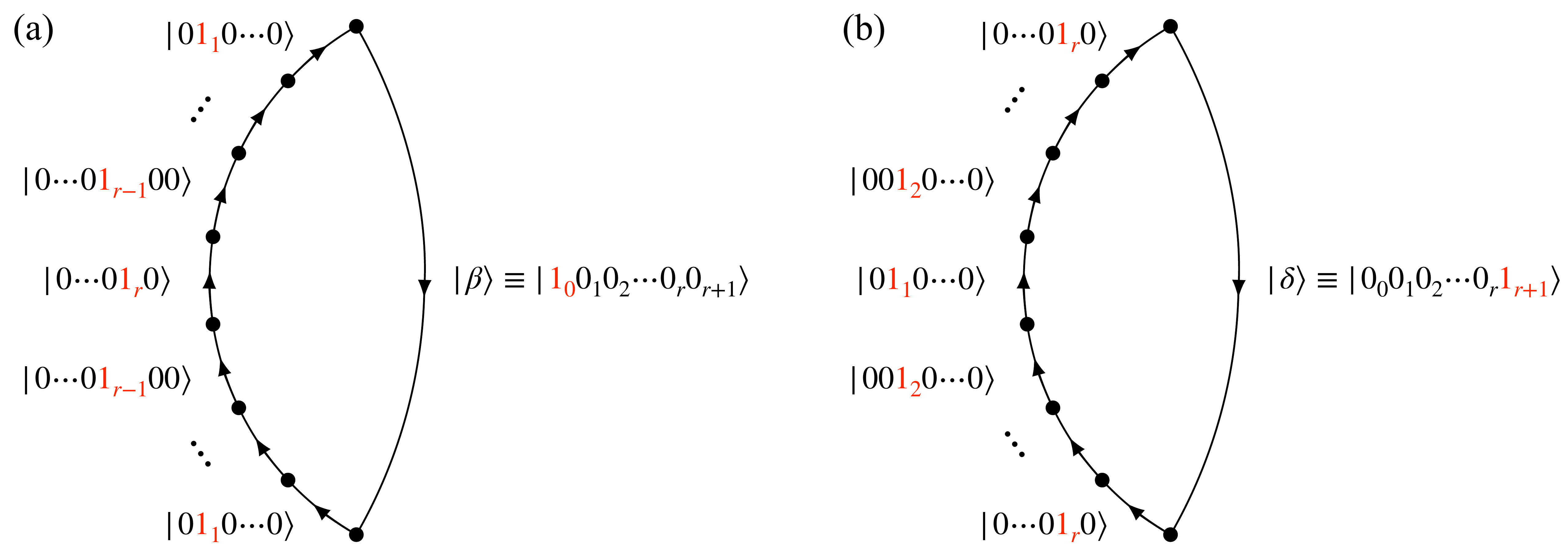}
    \caption{Diagrams of (a) $E^{(2r)}_\beta$ 
    and (b) $E^{(2r)}_\delta$ giving the nonzero lowest-order contribution to 
    $\Delta E_\alpha - \Delta E_\beta + \Delta E_\gamma - \Delta E_\delta$ 
    for the ferromagnetic state $|\alpha\rangle$ corresponding to the slowest scrambling.
    }
    \label{fig:sfig1}
\end{figure}

The ``one-hole'' diagrams~\cite{SzaboBook} shown in Fig.~\ref{fig:sfig1} represents 
the perturbation corrections $\Delta E_\beta$ and $\Delta E_\delta$ 
of the order $2r$, respectively. By following the recipe in Ref.~\cite{SzaboBook},
one can write down the evaluation results as
\begin{eqnarray}
    \Delta E_\beta^{(2r)} &=& 2 \left( \frac{J}{2}\right)^{2r} \frac{1}{A_1 A_2 \cdots A_{r-1} A_r A_{r-1} \cdots A_2 A_1}\,, \label{eq:Eb}\\
    \Delta E_\delta^{(2r)} &=& 2 \left( \frac{J}{2}\right)^{2r} \frac{1}{B_r B_{r-1} \cdots B_2 B_1 B_2 \cdots B_{r-1} B_r}\,, \label{eq:Ed}
\end{eqnarray}
where $A_i \equiv h_0 - h_i$ and $B_i \equiv h_{r+1} - h_i$.

\subsection{double excitation}

For the state $|\gamma\rangle = \hat{\sigma}^x_{r+1}\hat{\sigma}^x_0|\alpha\rangle = |1_0 0_1 0_2 \cdots 0_r 1_{r+1}\rangle$
with double excitation, there are many lowest-order diagrams 
contributing to $J_\mathrm{eff}^\mathrm{FM}$ as the two particles ($1$'s)
moving through the intervening region generates many different sequences
of intermediate states. These diagrams can be categorized 
into three types. Below we describe the evaluation of each type of diagram.

First, there are two one-hole diagrams for the special cases 
where one of the two particles is fixed and does not move. The diagrams are 
essentially the same as the ones given in Fig.~\ref{fig:sfig1}, 
while $0_{r+1}$ should be replaced with $1_{r+1}$ in Fig.~\ref{fig:sfig1}(a) 
and, $0_0$ should be replaced with $1_0$ in Fig.~\ref{fig:sfig1}(b).
By modifying Eqs. \eqref{eq:Eb} and \eqref{eq:Ed},
these two one-hole diagrams are evaluated straightforwardly as
\begin{eqnarray}
    2 \left( \frac{J}{2}\right)^{2r}
    \frac{1}{(A_r - J_z)\prod_{i=1}^{r-1} A_i^2}\,, \label{eq:EcA} \\
    2 \left( \frac{J}{2}\right)^{2r}
    \frac{1}{(B_1 - J_z)\prod_{i=2}^{r} B_i^2}\,, \label{eq:EcB}
\end{eqnarray}
where the interaction $J_z$ appears because the flow in the diagram goes 
through the ``interacting'' intermediate state where the particle from one end 
travels along the chain all the way to meet the fixed particle at the other end.

\begin{figure}
    \centering
    \includegraphics[width=0.95\textwidth]{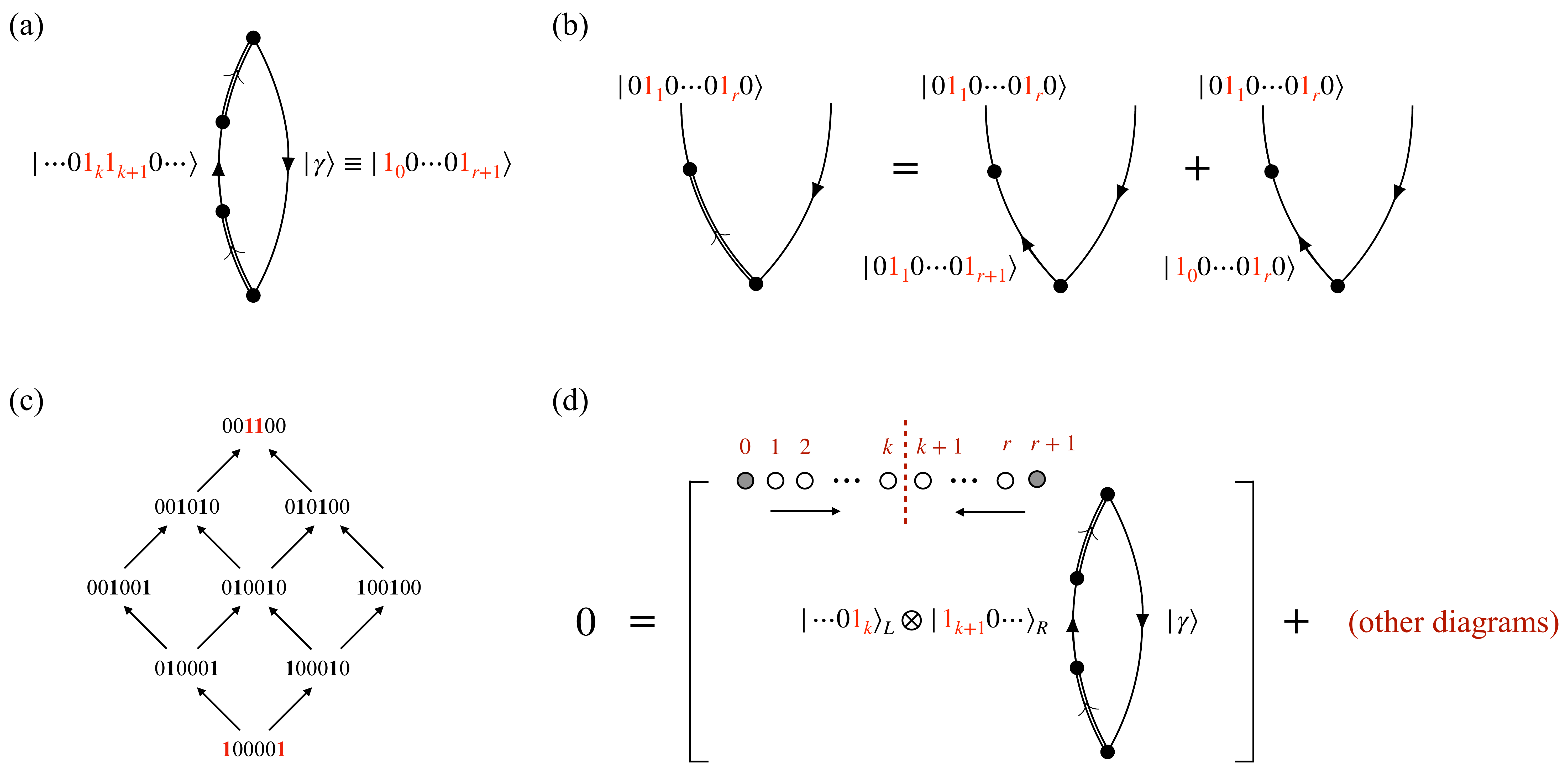}
    \caption{Diagrams for the lowest-order contributions of 
    $E_\gamma$ to $J_\mathrm{eff}^\mathrm{FM}$. 
    (a) Double-lined one-hole diagram with the interacting 
    intermediate state $|\cdots 1_k 1_{k+1} \cdots\rangle$. 
    (b) Example of the double line made of the addition of the two previous branches
    in the hierarchy. 
    (c) Example of the hierarchy to reach the intermediate state $|001100\rangle$
    from $|\gamma\rangle = |\cdots 100001 \cdots\rangle$. The left (right) arrows
    indicate the movement of the left (right) spin.
    (d) Evaluation of the sum of all the other diagrams that do
    not involve the interacting intermediate states.
    }
    \label{fig:sfig2}
\end{figure}

Second, there are more general cases where two particles move around
and meet together in the middle of the intervening region, generating
many one-hole diagrams with an intermediate state such as
$|0\cdots 0_{k-1} 1_k 1_{k+1} 0_{k+2} \cdots 0 \rangle$.
The ``particle'' line~\cite{SzaboBook} of those diagrams comes in three parts. 
In the lower part of the diagram, the two particles move towards
to the $k$ and $k+1$ sites, they meet at the intermediate state
in the middle part of the diagram, and then they depart from
each other and come back to the starting point $|\gamma\rangle$ in the upper part.
The complication in the evaluation of such diagrams is due to the fact
that one has to add all combinations of the sequences of 
the two-particle configurations, which we denote by the double lines
in Fig.~\ref{fig:sfig2}. 

The double line can be evaluated recursively. The summation of 
the partial diagrams are done along the tree-like hierarchy of 
the two-particle movements. Each intermediate state is reached through 
the two branches from the previous layer, unless one of the particles
still stays at the initial site, as each branch allows one particle
to move from either the left-hand side or the right-hand side. 
An example of the addition of the two branches
is shown in Fig.~\ref{fig:sfig2}(b), which occurs through
the hierarchical graph as exemplified in Fig.~\ref{fig:sfig2}(c). 
Summing up the partial diagrams at every vertex recursively 
along the graph, one can find that the movements of the left 
and right particles factorize after the summation, where a single
move creates the factor $1/A$ (left-particle) or $1/B$ (right-particle).
The double line in Fig.~\ref{fig:sfig2}(a) is then proportional to
\begin{equation*}
    \frac{A_k + B_{k+1}}{\prod_{i=1}^k A_i \prod_{i=k+1}^r B_i}.
\end{equation*}
The same expression can also be obtained for the upper double line 
as we start the summation from the upper end of the diagram.
Therefore, the diagram with the intermediate state 
$|\cdots 1_k 1_{k+1} \cdots \rangle$ in Fig.~\ref{fig:sfig2}(a) 
is written as
\begin{equation} \label{eq:EcK}
    2 \left( \frac{J}{2}\right)^{2r} \cdot 
    \frac{A_k + B_{k+1}}{\prod_{i=1}^k A_i \prod_{i=k+1}^r B_i}
    \cdot \frac{1}{A_k + B_{k+1} - J_z} \cdot
    \frac{A_k + B_{k+1}}{\prod_{i=1}^k A_i \prod_{i=k+1}^r B_i}\,.
\end{equation}

Last, we also need to take care of the other diagrams that do not visit
such an ``interacting" intermediate state $|\cdots 1_k 1_{k+1} \cdots\rangle$. 
These diagrams describe the cases where the two particles never come 
together to the neighboring sites and thus are independent of $J_z$.
Although the recursive summation described above cannot be used
for the direct summation of such non-interacting diagrams,  
the summation can still be evaluated using the following trick.

Let us prepare a system with the bond cut between the sites 
$k$ and $k+1$ as sketched in Fig.~\ref{fig:sfig2}(d), which 
separates the Hilbert space of the system into two parts of
the left and right chains.
The essential point of introducing this tweak is 
that the new system still produces the same non-interacting 
diagrams as the original interacting system does, 
while the sum over all the diagrams is just zero in the new system 
because the lack of coupling between $k$ and $k+1$ prohibits
the perturbation operators of the left and right regions appearing 
together in the operator string in the perturbation expansion. 
Therefore, we write down the diagrammatic equation shown 
in Fig.~\ref{fig:sfig2}(d) by using the evaluation of the diagram 
in Fig.~\ref{fig:sfig2}(a) in the setting of $J_z=0$ for its 
non-interacting cousin appearing as the first term on the right-hand side.
The second term ``(other diagrams)'' on the right-hand side is the non-interacting diagrams 
that we want to evaluate. 
The sum of the non-interacting diagrams is then given as
\begin{equation} \label{eq:EcNint}
    - 2 \left( \frac{J}{2}\right)^{2r} \cdot
    \frac{A_k + B_{k+1}}{\prod_{i=1}^k A_i^2 \prod_{i=k+1}^r B_i^2}.
\end{equation}
Adding up Eqs. (\ref{eq:EcA}-\ref{eq:EcNint}) for all $k$'s, we obtain 
the lowest-order expression of $\Delta E_\gamma$ as
\begin{equation*}
    \Delta E^{(2r)}_\gamma = 2 \left( \frac{J}{2}\right)^{2r}
    \left[
    \frac{1}{(A_r - J_z)\prod_{i=1}^{r-1} A_i^2}
    + \frac{1}{(B_1 - J_z)\prod_{i=2}^{r} B_i^2}
    + \sum_{k=1}^{r-1} \frac{A_k + B_{k+1}}{\prod_{i=1}^k A_i^2 \prod_{i=k+1}^r B_i^2} 
    \left( \frac{J_z}{A_k + B_{k+1} - J_z} \right) 
    \right].
\end{equation*}
Finally, we write $J_\mathrm{eff}^\mathrm{FM}(r) \equiv \Delta E_\alpha^{(2r)} - \Delta E_\beta^{(2r)} + \Delta E_\gamma^{(2r)} - \Delta E_\delta^{(2r)}$ as 
\begin{equation*}
    J_\mathrm{eff}^\mathrm{FM}(r) = 
    2 \left( \frac{J}{2}\right)^{2r} \cdot J_z \cdot 
    \Bigg[
    \frac{1}{A_r(A_r - J_z)\prod_{i=1}^{r-1} A_i^2}
    + \frac{1}{B_1(B_1 - J_z)\prod_{i=2}^{r} B_i^2}
    + \sum_{k=1}^{r-1}
    \frac{A_k + B_{k+1}}{(A_k + B_{k+1} - J_z)\prod_{i=1}^k A_i^2 \prod_{i=k+1}^r B_i^2}
    \Bigg].
\end{equation*}
Note that the formula of $J_\mathrm{eff}^\mathrm{FM}$ in the main text 
is given in the shorter form by defining an empty product as unity.